\documentclass[12pt,preprint]{aastex}

\begin{document}
\title{Branching ratio of Type Ib/c supernovae into GRB-supernovae}
\author{Maurice H.P.M. van Putten}
\affil{LIGO Project, NW 17-161, 175 Albany Street, Cambridge, MA 02139}
\begin{abstract}
We study centered and decentered nucleation of black holes in core-collapse of massive stars 
in binaries. By Bekenstein's gravitational-radiation recoil mechanism, a newly nucleated black 
hole typically leaves the central core prematurely. With low probability, the black hole 
remains centered and matures to a high-mass black hole which spins rapidly if the 
binary is compact. GRB030329/SN2003dh demonstrates that Type Ib/c supernovae 
are the parent population of long GRBs, whose branching ratio is 
${\cal R}=(2-4)\times 10^{-3}$. We identify ${\cal R}$ with the low probability of 
centered nucleation in compact binaries. Decentered events are predicted to produce 
a single short burst in gravitational radiation. Centered events are predicted to 
produce a second, long-burst in gravitational radiation powered by a luminous black hole. 
These signatures are of interest to LIGO, VIRGO and TAMA.
\end{abstract}

\section{Introduction}

GRB030329/SN2003dh \citep{sta03,hjo03} and 
GRB980425/SN1998bw \citep{gal98} show that Type Ib/c supernovae are the parent 
population of long GRBs. Type Ib/c SNe are believed to represent core-collapse 
events of massive stars in compact binaries \citep{woo93,pac98,bro00,bet03}. They 
are probably part of a continuous sequence adjacent to Type II SNe, 
ordered by increasing compactness of the binary in which the hydrogen (Ib/c) and 
the helium (Ic) envelope are removed in a common envelope phase \citep{nom95,tur03a}. 
The remaining naked star rotates rapidly at the orbital period by tidal spin-up. As 
the inactive iron-core succumbs to its own weight and that of the surrounding 
He-envelope, a rotating black hole nucleates during core-collapse \citep{bet03}. 
Some of the binding energy liberated during gravitational collapse will be channeled 
to eject matter, producing an accompanying hydrogen (and helium) deficient Type Ib 
(Type Ic) supernova \citep{mac03}.

The branching ratio of Type Ib/c SNe to GRB-SNe can be calculated from the ratio 
$(1-2)\times 10^{-6}$ of observed GRBs-to-Type II supernovae \citep{por01}, a beaming 
factor of 450 \citep{mvp03b} to 500 \citep{fra01} and a rate of about 0.2 of Type 
Ib/c-to-Type II supernovae \citep{tur03b}, giving
\begin{eqnarray}
{\cal R}[\mbox{Ib/c}\rightarrow\mbox{GRB}]=\frac{N(\mbox{GRB-SNe})}{N(\mbox{Type~Ib/c})}
\simeq (2-4)\times 10^{-3}.
\label{BRANCH}
\end{eqnarray}
This ratio is remarkably small, suggesting a higher-order down-selection process.

The small branching ratio (\ref{BRANCH}) can be attributed to various factors in the process
of creating GRBs in Type Ib/c supernovae \citep{pod04}, e.g, not all baryon poor jets 
successfully punch through the remnant stellar envelope \citep{mac99}, and not all massive progenitors 
making Type Ib/c supernovae nucleate rapidly rotating black holes. It is unlikely that 
either one of these down-selection processes by itself accounts for the smallness of ${\cal R}$.
Rather, a combination of these might effectively contribute to a small branching ratio. 

By tidal interaction with the companion star, the naked star is not spherical prior to collapse. 
Black holes nucleated in nonspherical collapse possess recoil by Bekenstein's gravitational
radiation recoil mechanism \citep{bek73}. Tidal deformation produces a systematic recoil 
velocity, which may combine with random multipole mass-moments to produce a distribution in 
recoil velocities. Some of the black holes will leave the central high-density core prematurely, 
before completion of the stellar collapse process. These events are {\em decentered}. Other
holes will remain centered and surge into a high-mass object surrounded by a high-density accretion 
disk or torus. These events are {\em centered}. Centered black holes becomes luminous in a state of 
suspended accretion, if they rotate rapidly. They spin down against emissions in gravitational 
radiation and other radiation channels \citep{mvp03a}. The latter includes a burst in high-energy 
radiation from torus winds which radiatively drives a supernova \citep{mvp04}, closely related to 
\citep{bet03,lee02}.

Here, we quantify the various stages in the nucleation of black holes in stellar collapse. We 
favor an association with binaries \citep{nom95,tur03a} based on the Type II/Ib event SN1993J 
\citep{mau04} and the proposed association of GRB-supernovae remnants with soft X-ray transients 
\citep{bet03}. We shall identify a branching ratio of core-collapse events producing centered 
nucleation of black holes with the probability of low kick velocities based on the Bekenstein
recoil mechanism.

A related but different mechanism for explaining the small branching ratio based on kick 
velocities in core-collapse poses fragmentation into two or more objects \citep{dav02}. In
this scenario, GRBs are associated with the formation of a fireball in the merger of 
binaries possessing small kick velocities. It is motivated, in part,
in the search for delay mechanisms in creating a GRB, after the onset of the supernova on 
the basis of X-ray line-emissions in GRB011211. However, X-ray line-emissions produced in
radiatively powered supernovae allow the same time-of-onset of the GRB and the supernova,
obviating the need for any delay mechanism \citep{mvp04}. 

\section{Centered Nucleation}

Rotating black holes are described by Kerr \citep{ker63}. In core-collapse of massive stars, 
rotating black holes nucleate by accumulation of mass and angular momentum from infalling 
matter. The Kerr solution describes the constraint
\begin{eqnarray}
J_H\le GM^2/c
\label{EQN_JM}
\end{eqnarray}
for a black hole of mass $M$ and angular momentum $J_H$, where $G$ is Newton's constant and 
$c$ is the velocity of light. Table I summarizes the key quantities of Kerr black holes.
Quite generally, initial collapse of a rotating core produces a torus \citep{ree74,due04}, 
which initially satisfies $J_T>GM_T^2/c.$
Thus, the nucleation of black holes takes place through a {\em first-order} phase-transition: a 
torus forms of increasing mass by accumulation of matter, diluting its angular momentum until it 
satisfies (\ref{EQN_JM}) and collapses into an extremal black hole. The alternative of a 
second-order phase transition which initially forms a sub-solar mass black hole, requires rapid 
shedding of excess angular momentum by gravitational radiation. However, limited mass-densities 
in core-collapse probably render this mechanism ineffective in competition with mixing on the 
free-fall timescale of the core. Nevertheless, gravitational radiation emitted from a 
non-axisymmetric torus prior to the nucleation of the black hole is potentially interesting 
\citep{ree74,due04}.

Gravitational radiation in the formation of black holes through a first-order phase transition is 
important in non-spherical collapse, even when its energy emissions are small relative to the 
initial mass of the black hole. The Bekenstein gravitational radiation-recoil mechanism operates 
already in the presence of initial asphericities of about $10^{-3}$, producing a recoil
of 300km/s or less. The radius of the accretion disk or torus around a newly formed stellar 
mass black hole is $R_T\sim 10^7$cm. A torus of a few tenths of a solar mass forms by 
accumulation of matter spiralling in, compressed by a factor $\sim(r/r_{ISCO})^4$ as it 
stalls against the angular momentum barrier outside the inner most stable circular orbit 
(ISCO). The time-of-collapse of stellar matter from a radius $r$ is about
the free-fall timescale, 
\begin{eqnarray}
t_{ff}\simeq30\mbox{~s}~\left(\frac{M_{He}}{10M_\odot}\right)^{-1/2}
              \left(\frac{r}{10^{10}\mbox{cm}}\right)^{3/2},
\end{eqnarray}
where $M_{He}$ denotes the mass of the progenitor He-star.
The newly formed low-mass black hole is typically kicked out of the central 
high-density region into surrounding lower-density regions {\em before} core-collapse is 
completed. It continues to grow off-center by accretion of relatively 
low-density matter -- a high-density accretion disk never forms. With low but non-zero 
probability, the black hole has small recoil, remains centered and surges 
into a high mass black hole surrounded by a high-density torus. 

After nucleation of the black hole, an accretion disk may form provided that the specific 
angular momentum $j_m$ of infalling matter exceeds that of the inner most stable circular 
orbit (ISCO),
\begin{eqnarray}
j_m \ge l(a/M) GM/c
\label{EQN_JL}
\end{eqnarray}
on the angular momentum $J_H$ of a black hole of mass $M$. Here, $l(a/M)$ 
denotes the dimensionless specific angular momentum of matter in circular 
orbits on the ISCO, where $a=J_H/M$ denotes the specific angular momentum 
of the black hole. Explicitly, we have \citep{bar70,bar72}
$l=({2}/3\sqrt{3})\left(1+2\sqrt{3z-2}\right)$ in terms of 
$z={r_{ISCO}}/{M}=3+Z_2-\left[(3-Z_1)(3+Z_1+2Z_2)\right]^{1/2}$ with
$Z_1=1+(1-q)^{1/3}\left[ (1+q)^{1/3} + (1-q)^{1/3} \right],~~
Z_2=\left(3q^2+Z_1^2\right)^{1/2}$ and $q=a/M$.
Notice that $2/\sqrt{3}\le l \le 2\sqrt{3}$, between an extremal black hole 
$(a=M,~z=1$) and a non-rotating black hole $(a=0,~z=6$). The evolution of 
the newly nucleated black hole continues to be governed by angular momentum 
loss of the surrounding matter, until the inequality in (\ref{EQN_JL}) is 
reversed.

The black hole rapidly grows uninhibited, while the inequality (\ref{EQN_JL}) is reversed
($j_m < l(a/M) GM/c)$. This {\em surge} continues until 
once again (\ref{EQN_JL}) holds. In dimensionless form, (\ref{EQN_JL}) becomes
\begin{eqnarray}
l\left(\frac{\beta j(s)}{m^2(s)}\right)=\frac{k_1\beta s^2}{m(s)},
\end{eqnarray}
where
\begin{eqnarray}
\beta=k_2\frac{\omega c s_0}{GR\rho_c}=4.22 k_2P_d^{-1}R_1^{-1}(M_{He}/10M_\odot)^{-1/3}
\label{EQN_NORM}
\end{eqnarray}
in terms of the dimensionless integrals $j(s)=4\pi\int^s_0 \hat{\rho} s^4 ds$ and
$m(s)=4\pi\int_0^s \hat{\rho} s^2 ds$ of the normalized Lane-Emden density 
distribution with $\hat{\rho}=1$ at the origin and the zero $\hat{\rho}=0$ at 
$s_0=6.89685$. Here, $(k_1,k_2)=(1,1)$ in cylindrical geometry for which
$j_m=\omega r^2$, and $(k_1,k_2)=(5/3,2/3)$ in spherical geometry for which
$j_m=(2/3)\omega r^2$; $P_d$ denotes the binary period in days, $R_1$ denotes
the radius in units of the solar radius $6.96\times 10^{10}$cm \citep{kip90},
and $M_{He}$ the mass of the progenitor He-star.

Fig. 1 shows the solutions as a function of dimensionless period $1/\beta$.
The upper branch shows that rapidly spinning black holes plus accretion disk
form in small-period binaries \citep{bet03,lee02}, 
following a surge for periods beyond the bifurcation points
\begin{eqnarray}
\mbox{cylindrical geometry}~ (\beta=6.461):~~\frac{a}{M}=0.9541,
~~\frac{E_{rot}}{E_{rot}^{max}}=0.6624,~~\frac{M}{M_{He}}=0.4051.
\label{EQN_AA1}
\end{eqnarray}
\begin{eqnarray}
\mbox{spherical geometry}~ (\beta=5.157):~~\frac{a}{M}=0.7679,
~~\frac{E_{rot}}{E_{rot}^{max}}=0.3220,~~\frac{M}{M_{He}}=0.3554.
\label{EQN_AA2}
\end{eqnarray}
The resulting mass and energy fractions as a function of $1/\beta$ are shown in
Fig. 2. These two geometries serve to bound the range of values in more
detailed calculations, e.g., through multi-dimensional numerical
simulations. 

{\em The Bardeen trajectory} corresponds to continuing accretion beyond surge, 
wherein matter remaining in the remnant envelope forms an
accretion disk outside the ISCO. At this point, magnetohydrodynamical 
stresses within the disk as well as disk winds may drive continuing 
accretion. Accretion from the ISCO onto the black hole further increases the black 
hole mass and spin according to $zM^2=\mbox{const.}$ \citep{bar70},
generally causing spin-up towards an extremal state of the black hole.
In Fig. 1 this is indicated by accretion {\em upwards} 
beyond the upper ISCO-branch.

{\em Radiative spin-down} corresponds to a long-duration burst of gravitational
radiation emitted by a non-axisymmetric torus \citep{mvp99,mvp04}, described by
a frequency and energy
\begin{eqnarray}
E_{gw}=(4\times 10^{53} \mbox{~erg}) M_{7} \eta_{0.1}
       \left(\frac{E_{rot}}{E_{rot}^{max}}\right),~~
f_{gw}=(500\mbox{~Hz})M_7\eta_{0.1},
\label{EQN_GW}
\end{eqnarray}
where $M_7=M/7M_\odot$, and $\mu=M_T/0.03M$ and $\eta=\Omega_T/\Omega_H$ denote the relative 
mass and angular velocity of the torus. 
This takes place if the torus is uniformly magnetized with the remnant magnetic field of the 
progenitor star. In Fig. 1, this radiative spin-down is indicated by 
a transition {\em downwards} from the upper ISCO branch to the branch on which the angular 
velocities of the black hole and of matter at the ISCO match ($\Omega_H=\Omega_{ISCO}$ and 
$\eta=1$). This radiative transition lasts for the lifetime of rapid spin of the
black hole -- a dissipative timescale of tens of seconds \citep{mvp03a}. 
Additional matter accreted is either blown off the torus in its winds, or accumulates 
and accretes onto the black hole after spin-down.

\section{Branching ratio}

In what follows, we shall consider a two-dimensional 
Gaussian distribution of kick velocities in the equatorial plane associated with the the 
tidal deformation of the progenitor star by its companion, assuming a velocity dispersion 
$\sigma_{kick}\simeq 100$km/s in Bekenstein's recoil mechanism.

The probability of centered nucleation during $t_{ff}\simeq 30$ s is that of a kick velocity
$v_{kick}<v^*=$10km/s, i.e.:
\begin{eqnarray}
P_{c}=P(v_{kick}<v^*)\simeq 0.5\% \left(\frac{v^*}{10\mbox{km/s}}\right)^2
       \left(\frac{\sigma_{kick}}{100\mbox{km/s}}\right)^{-2}.
\end{eqnarray}
While the numerical value has some uncertainties, the selection mechanism by gravitational 
radiation-recoil effectively creates a small probability of centered nucleation.
We identify the branching ratio of Type Ib/c SNe into GRBs with the probability of centered
nucleation, 
\begin{eqnarray}
{\cal R}[\mbox{Ib/c}\rightarrow\mbox{GRB}]=P_{c}\simeq 0.5\%,
\end{eqnarray}
effectively creating a small, higher-order branching ratio. 

\section{Single and double bursts in gravitational radiation}

The proposed centered and decentered core-collapse events predicts a differentiation in 
gravitational wave-signatures. These signatures are of interest to the newly commissioned 
gravitational wave-detectors LIGO, VIRGO and TAMA, both as burst sources and through their 
collective contributions to the stochastic background in gravitational radiation \citep{mvp03a}. 

The black hole nucleation process is accompanied by a short burst in gravitational radiation,
specifically in response to non-axisymmetric toroidal structures and fragmentation 
\citep{bek73,ree74,dav02,pop04,due04}. Its gravitational radiation signature 
depends on details of the hydrodynamical collapse.
Centered nucleation is followed by a long burst in gravitational radiation.

A short burst in gravitational radiation is hereby common to all Type Ib/c supernovae. This 
may apply to Type II events as well. Type II events are possibly associated with low 
spin-rates and could represent delayed core-collapse via an intermediate ``nucleon" star 
(SN1987A, e.g. \cite{bet03}). Their gravitational wave-emissions are thereby essentially 
limited to that produced by kick (if any) and collapse of this nucleon star.  For a 
recent review of the short-duration ($<<1$s) bursts of gravitational waves in core-bounce, 
more closely related to Type II supernovae, see \citep{fry04} and references therein.

On account of (\ref{EQN_AA1}-\ref{EQN_AA2}), core-collapse compact can produce high-mass,
rapidly spinning black holes in centered nucleation, whose rotational energy can reach about
one-half the maximal spin-energy of a Kerr black hole. In a suspended accretion state, these
black holes spin-down in the process of emitting a long-duration burst of tens of seconds in
gravitational waves \citep{mvp04}. This long duration burst comes as a second burst, after the
short burst of gravitational radiation in centered nucleation.

To conclude, Type Ib/c supernovae produce a short, single burst of gravitational
radiation at birth of a low-mass black hole. The sub-population of GRB-supernovae produce a 
subsequent long burst of gravitational radiation representing spin-down of the black hole.
The second burst takes place after a quiescent or subluminous \citep{min02} surge of the black 
hole into a high-mass object. 

{\bf Acknowledgement.}
The author thanks P. Saulson, J. Bekenstein and the referee for constructive comments. 
    This research was
    completed at a joint APCTP-TPI meeting at the University of Alberta, and is
    supported by the LIGO Observatories, constructed by Caltech and MIT
    with funding from NSF under cooperative agreement PHY 9210038.
    The LIGO Laboratory operates under cooperative agreement
    PHY-0107417. This paper has been assigned LIGO document number
    LIGO-P040014-00-R.

\mbox{}\\

\newpage

\begin{table}
\begin{center}
\caption{Trigonometric parametrization of a Kerr black hole. Here, $M$ denotes
the mass of the black hole, $a=J_H/M$ denotes the specific angular momentum,
$E_{rot}$ the rotational energy and $M_{irr}$ denotes the irreducible mass.}
\mbox{~}\\
\begin{tabular}[t]{ll@{\quad}c@{\quad}c@{\quad}}
 {\sc Symbol} &  {\sc Expression} & {\sc Comment} \\
 \hline\\
 $\lambda$    &  $\sin\lambda = a/M$  & \\
 $\Omega_H$   &  $\tan(\lambda/2)/2M$ & \\
 $J_H$        &  $M^2\sin\lambda$  & \\
 $E_{rot}$    &  $2M\sin^2(\lambda/4)$& $\le 0.29 M$\\
 $M_{irr}$    &  $M\cos(\lambda/2)$   & $\ge 0.71 M$\\
 \mbox{}\\\hline
\end{tabular}
\end{center}
\end{table}

\begin{figure}
\plotone{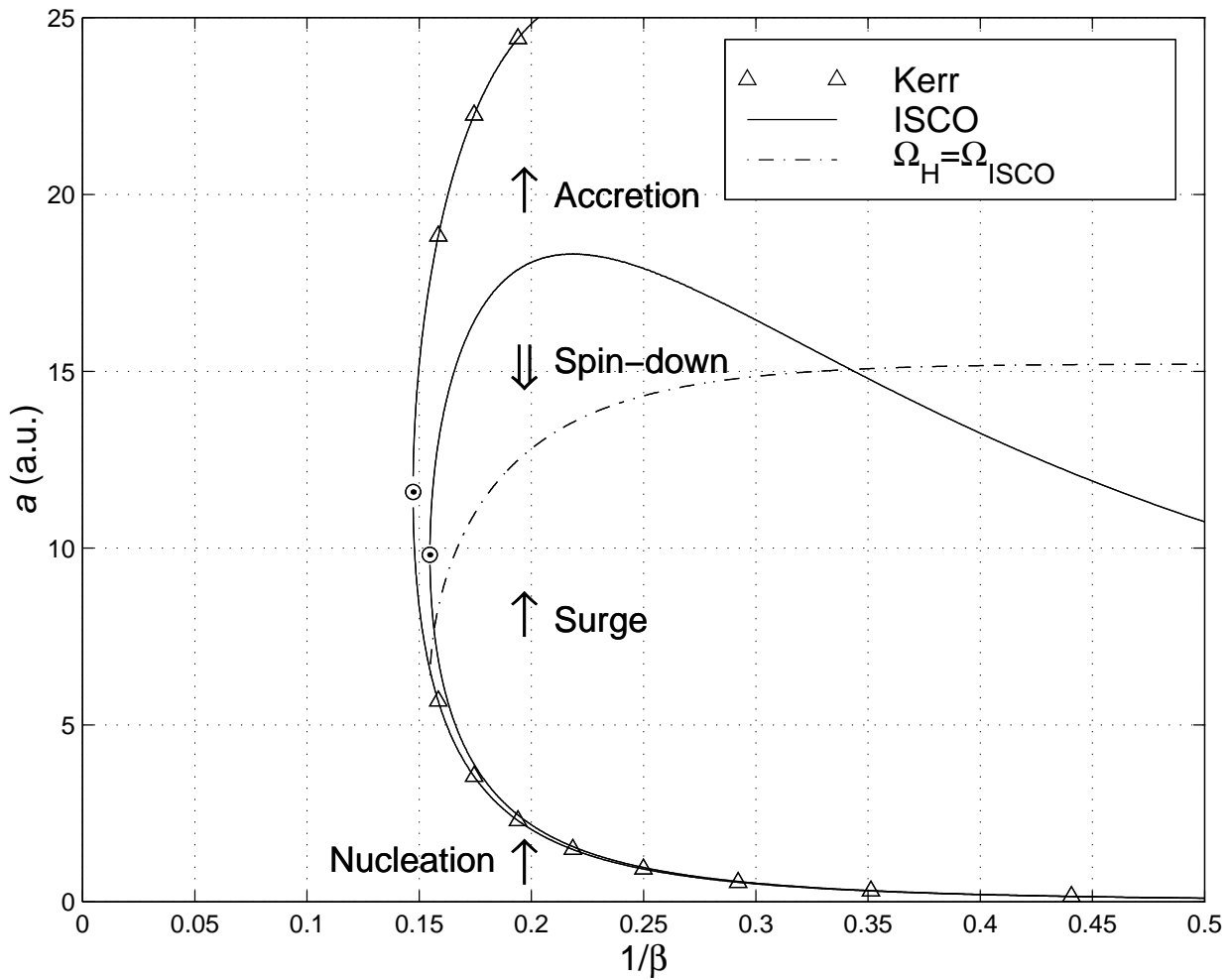}
\caption{Centered nucleation of black holes in core-collapse of a uniformly rotating
massive star: accumulated specific angular momentum of the central object (arbitrary units) 
versus dimensionless orbital period $1/\beta$. Arrows indicate the evolution as a function
of time. Kerr black holes exist {\em inside} the outer curve (diamonds). A black hole
nucleates following the formation and collapse of a torus, producing a short burst in
gravitational radiation. In centered nucleation, the black hole surges to a high-mass
object by direct infall of matter with relatively low specific angular momentum, up to
the inner continuous curve (ISCO). At this point, the black hole either spins up by
continuing accretion or spins down radiatively against gravitational radiation emitted
by a surrounding non-axisymmetric torus. In this state, the black hole creates a baryon 
poor jet along an open ``ergotube" as input to GRB-afterglow emissions. This state proceeds 
until the angular velocity of the black hole equals that of the torus (dot-dashed line). 
These curves are computed for a Lane-Emden mass-distribution with polytropic index $n=3$ 
in the limit of conservative collapse, neglecting energy and angular momentum loss in 
radiation and winds. Shown are the results in cylindrical geometry. This scenario fails 
in decentered collapse, where the newly nucleated black hole leaves the high-density core 
prematurely, prohibiting the formation of a high mass black hole surrounded by a 
high-density torus in a state of suspended accretion. The probability of centered 
nucleation defines the branching ratio of Type Ib/c supernovae into long GRBs.} 
\end{figure}
\begin{figure}
\plotone{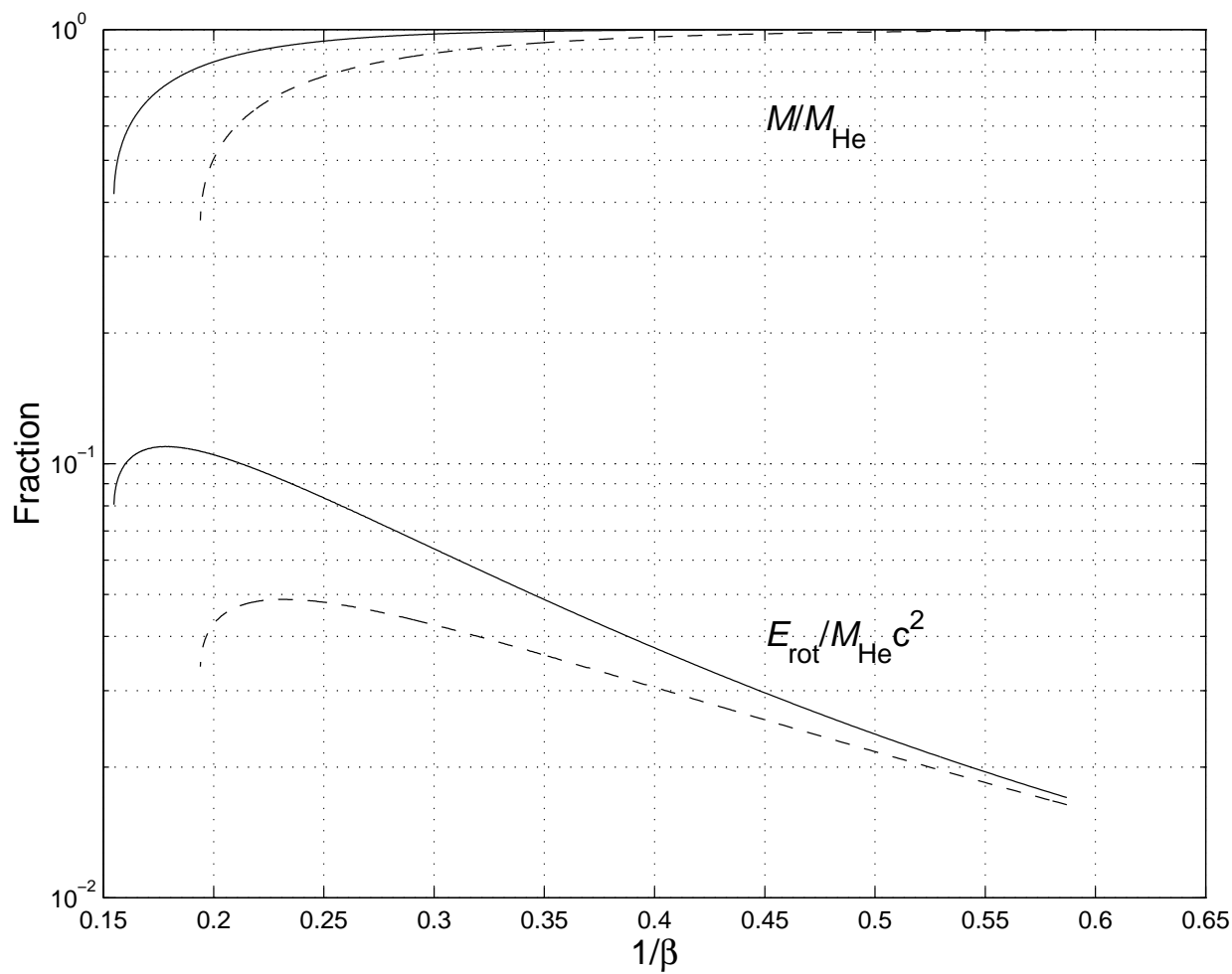}
\caption{The black hole mass $M$ and rotational energy $E_{rot}$ of formed after surge
in centered nucleation, expressed relative to the mass $M_{He}$ of the progenitor He-star. 
The results are shown in cylindrical geometry (continuous) and spherical geometric (dashed). 
Note the broad distribution of high-mass black holes with large rotational
energies of $5-10\%$ (spherical to cylindrical) of
$M_{He}c^2$.}
\end{figure}


\begin{thebibliography}{}
\bibitem[Bardeen(1970)]{bar70} Bardeen, J.M., 1970, Nature, 226, 64
\bibitem[Bardeen, Press \& Teukolsky(1972)]{bar72} Bardeen J.M., Press W.H., Teukolsky S.A., 1972, 178, 347
\bibitem[Bekenstein(1973)]{bek73} Bekenstein, J., 1973, ApJ, 183, 657
\bibitem[Bethe, Brown \& Lee(2003)]{bet03} Bethe, H.A., Brown, G.E., \& Lee, C.-H., 2003, Selected Papers: 
   Formation and Evolution of Black Holes in the Galaxy (World Scientific)
\bibitem[Brown et al.(2000)]{bro00} Brown, G.E., Lee, C.-H., Wijers R.A.M.J., Lee, H.K., Israelian G. 
 \& Bethe H.A., 2000, NewA, 5, 191
\bibitem[Cappellaro, Barbon \& Turatto(2003)]{tur03b} Cappellaro, E., Barbon, R., \& Turatto, M., 2003, astro-ph/0310859
\bibitem[Davies et al.(2002)]{dav02} Davies, M.B., King, A., Rosswog, S., \& Wynn, G., 2002, ApJ, 579, L63 
\bibitem[Duez, Shapiro \& Yo(2004)]{due04} Duez, M.D., Shapiro, S.L., \& Yo, H.-J., 2004, gr-qc/0401076
\bibitem[Frail et al.(2001)]{fra01} Frail, D.A., et al., 2001, ApJ, 562, L55
\bibitem[Fryer, Holz \& Hughes(2004)]{fry04} Fryer, C.L., Holz, D.E., \& Hughes, S.A., 2004, astro-ph/0403188
\bibitem[Galama et al.(1998)]{gal98} Galama, T.J., et al. 1998, Nature, 395, 670
\bibitem[Ghisellini et al.(2002)]{ghi02} Ghisellini, G., Lazatti, D., Rossi, E., \& Rees, M.J., 2002, A\&A, 389, L33
\bibitem[Hjorth et al.(2003)]{hjo03} Hjorth, J., et al., 2003, Nature, 423, 847
\bibitem[Hoeflich, Wheeler \& Wang(1999)]{hoe99} H\"oflich, P.J., Wheeler, J.C., \& Wang, L., 1999, ApJ, 521, 179
\bibitem[Kerr(1963)]{ker63} Kerr, R.P., 1963, Phys. Rev. Lett., 11, 237
\bibitem[Kippenhahn \& Weigert(1990)]{kip90} Kippenhahn, R. \& Weigert, A.
  1990, Stellar structure and evolution (Springer-Verlag: New York), p178
\bibitem[Kouveliotou et al.(1993)]{kou93} Kouveliotou, C., et al., 1993, ApJ, 413, L101
\bibitem[Lee, Brown \& Wijers(2002)]{lee02} Lee, C.-H., Brown, G.E., \& Wijers, R.A.M.J., 2002, ApJ, 575, 996
\bibitem[MacFadyen \& Woosley(1999)]{mac99} MacFadyen, A.I. and Woosley, S.E., 1999, ApJ, 524, 262
\bibitem[MacFadyen(2003)]{mac03} MacFadyen, A.I., 2003, astro-ph/0301425
\bibitem[Maund et al.(2004)]{mau04} Maund, J.R., Smartt, S.J., Kudritzki, R.P., Podsiadlowski, P.,
 \& Gilmore, G.F., 2004, Nature, 427, 129
\bibitem[Mineshige(2002)]{min02} Mineshige, S., Hosokawa, T., Machida, M., \& Matsumoto, R., 2002, PASJ, 54, 655
\bibitem[Nomoto, Iwamoto \& Suzuki(1995)]{nom95} Nomoto, K., Iwamoto, K., \& Suzuki, T, 1995, Phys. Rep. 256, 173
\bibitem[Paczynski(1998)]{pac98} Paczy\'nski, B., 1998, ApJ, 494, L45
\bibitem[Podsiadlowski et al.(2004)]{pod04} Podsiadlowski, Ph., Mazzali, P.A., Nomoto, K., Lazzati, D.,
 \& Cappellaro, E., 2004, ApJ, 607, L17
\bibitem[Popov(2004)]{pop04} Popov, S.B., 2004, astro-ph/0403710
\bibitem[Porciani \& Madau(2001)]{por01} Porciani, C., \& Madau, P., 2001, ApJ, 548, 522
\bibitem[Reeves et al.(2002)]{ree02} Reeves, J.N., et al., 2002, Nature, 416, 512
\bibitem[Turatto(2003)]{tur03a} Turatto, M., 2003, in Supernovae and Gamma-ray Bursters, ed.
    K.W. Weiler (Springer-Verlag, Heidelberg), p21
\bibitem[van Putten \& Levinson(2003)]{mvp03a} van Putten M.H.P.M., \& Levinson, A., 2003, ApJ, 584, 937
\bibitem[van Putten et al.(2004)]{mvp04} van Putten M.H.P.M., Levinson, A., Regimbau, T., Punturo, M., \& Herry, 
    G.M., 2004, Phys. Rev. D. 69, 044007
\bibitem[van Putten(1999)]{mvp99} van Putten, M.H.P.M., 1999, Science, 284, 115
\bibitem[van Putten \& Regimbau(2003)]{mvp03b} van Putten, M.H.P.M., \& Regimbau, T., 2003, ApJ, 593, L15
\bibitem[Rees, Ruffini \& Wheeler(1974)]{ree74} Rees, M.J., Ruffini, R., Wheeler, J.A., 1974, {\em Black holes, Gravitational 
   waves and Cosmology: an introduction to current research} (Gordon \& Breach, 
   New York), Section 7
\bibitem[Stanek et al.(2003)]{sta03} Stanek, K.Z., et al. 2003, ApJ, 591, L17
\bibitem[Woosley(1993)]{woo93} Woosley, S.E., 1993, ApJ, 405, 273

\end{thebibliography}
\end{document}